\documentclass[12pt]{article}
\usepackage{amsmath,amssymb,mathrsfs,hyperref,slashed,bbm}
\usepackage{tensor}
\usepackage{cleveref}
\crefname{equation}{Eq.}{Eqs.}
\usepackage[square, numbers, comma, sort&compress]{natbib}

\newcommand{\avg}[1]{\left< #1 \right>} 

\def\approxless{%
  \def\p{%
    \setbox0=\vbox{\hbox{$<$}}%
    \ht0=0.6ex \box0 }%
  \def\s{%
    \vbox{\hbox{$\sim$}}%
  }%
  \mathrel{\raisebox{1ex}{%
      \mbox{$\underset{\s}{\p}$}%
    }}%
}

\begin{document}
\title{\vspace{-1.8in}
{Graviton multi-point amplitudes for higher-derivative gravity in anti-de Sitter space}}
\author{\large M.M.W. Shawa$^{(1)}$,  A.J.M. Medved$^{(1,2)}$
\\
\vspace{-.5in} \hspace{-1.5in} \vbox{
 \begin{flushleft}
$^{\textrm{\normalsize (1)\ Department of Physics \& Electronics, Rhodes University,
  Grahamstown 6140, South Africa}}$
$^{\textrm{\normalsize (2)\ National Institute for Theoretical Physics (NITheP), Western Cape 7602,
South Africa}}$
\\ \small \hspace{1.07in}
  markshawa@aims.ac.za,\  j.medved@ru.ac.za
\end{flushleft}
}}
\date{}
\maketitle
\abstract{
We calculate graviton multi-point amplitudes in an anti-de Sitter black brane background for higher-derivative gravity of arbitrary order in numbers of derivatives. The calculations are performed using tensor graviton modes in a particular regime of comparatively high energies and large scattering angles. The regime simplifies the calculations but, at the same time, is well suited for translating these results into the language of the dually related gauge theory. After considering theories of up to eight derivatives, we generalize to even  higher-derivative theories by constructing a ``basis'' for the relevant scattering amplitudes. This construction enables one to find the basic form of the $n$-point amplitude for arbitrary $n$ and any number of derivatives. Additionally, using the four-point amplitudes for six and eight-derivative gravity, we re-express the scattering properties in terms of the Mandelstam variables.
}

\section{Introduction}\label{intro}
The gauge--gravity duality enables one to describe a $d$-dimensional gauge theory in terms of a $(d+1)$-dimensional gravitational theory~\cite{Maldacena}. Importantly, the duality relates  a strongly coupled field theory to
 a weakly coupled theory of gravity.  Since strongly coupled gauge theories are not very well understood, the duality provides a means for making analytical statements about them. One  application of this framework \cite{Hofman} is the correspondence between stress--energy tensor correlation functions in the relevant gauge theory 
and graviton
scattering amplitudes in its gravitational dual \cite{Witten}.

In the earliest investigations into the duality --- which mostly focused on  5-dimensional anti-de Sitter (AdS) space and 4-dimensional super Yang--Mills theory --- the rank of the gauge theory $N$ is taken to infinity, which then corresponds to Einstein's theory of gravity \cite{Aharony}. (On the other hand, the 't Hooft coupling $\lambda = g^2_s N$ is regarded as large but finite in the standard limit.) With deviations to large but finite values of $N$, the gravitational dual
can be expected  to include higher-derivative corrections in addition to Einstein's (two-derivative) Lagrangian~\cite{Cheung}. If the interest is only in  gauge-invariant quantities (such as scattering amplitudes), then
one  can limit considerations to the multi-derivative terms in the Lagrangian which are strictly composed of contractions between  ``proper'' four-index Riemann tensors ({\em i.e.}, two-index tensors and scalars are excluded).  This simplification was recently discussed in~\cite{Deser} and can be shown through a gauge transformation of the graviton  that involves the Ricci scalar and tensor~\cite{tHooft2,Pollock2}. 

In this paper, we calculate graviton scattering amplitudes for higher-derivative theories of gravity in an AdS black brane background. Our approach is similar to that of~\cite{Brustein1}, where the focus is on Einstein and four-derivative gravity. However, those scattering amplitudes could be substantially simplified by enforcing on-shell conditions. In the case of Einstein's gravity, 
the equations of motion can be used to eliminate amplitudes with two derivatives acting on single graviton. Meanwhile, four-derivative gravity can always be gauge transformed to a Gauss--Bonnet theory  by using the ``inverse" of the aforementioned transformation. As a Lovelock extension of Einstein's theory, Gauss--Bonnet gravity has equations of motion that have at most two derivatives. Thus, the very same logic and simplification  applies in this case as well.
 (See the Appendix in~\cite{Brustein1} for further details.) In the current work, we do not have the luxury of restricting to Lovelock theories or theories
that are related to  Lovelock   via a  gauge transformation;~\footnote{This follows from any $2k$-derivative Lovelock extension vanishing identically in $2k-1$ or less dimensions.} meaning that the previous simplification can no longer be relied upon.  It follows that terms with two derivatives acting on a graviton must now be included in the calculation of amplitudes. 

Starting with six and eight-derivative theories in Section~2, we utilize what could be called ``basis amplitudes" to construct the $n$-point amplitudes for arbitrarily large $n$. The generalization of these basis amplitudes mitigates the task of finding multi-point amplitudes for theories with an arbitrary number of derivatives (see~\Cref{section-riemqgravity}). Finally, using $4$-point amplitudes in Section~4, we re-express the scattering properties of the six and eight-derivative theories in terms of the Mandelstam variables. (Section~5 contains
a brief conclusion.)

Our intention is, in a later paper, to map the current results to stress--tensor correlation functions in the dual gauge theory. The motivation for such a 
treatment  is
to learn about the gravitational dual to the gauge theory describing the quark--gluon plasma and other strongly coupled gauge theories~\cite{CasalderreySolana}. These findings could conceivably be tested  thanks to the  high-energy hadron-collider  laboratories  at Brookhaven and CERN. 

Because our long-term goal is to apply the  gauge--gravity duality for
the purpose of making experimentally viable predictions, 
we are mainly  interested in  the boundary limit of the amplitudes and, moreover, only those contributions that would survive holographic renormalization
\cite{Skenderis20} and
could be  discernible in gauge-theory correlation functions.
With this in mind, our calculations are limited to a special kinematic region which was  referred to as the ``high-momentum regime" in~\cite{Brustein1}. As discussed later, this regime is particularly well suited for discriminating the higher-derivative contributions in the graviton scattering amplitudes and then, ultimately, in the dual correlation functions. 
    
\subsection{Summary of formalism}

Much of groundwork for the following analysis has already been laid out in~\cite{Brustein1,Brustein2}. Here, we will summarize some of the key elements that are necessary for the task at hand. 

The AdS black brane background in a five-dimensional spacetime~\footnote{Generalizations to other dimensionalities are straightforward.} has a metric of
the form
\begin{align}\label{ads-metric}
		ds^2\; =\; -f(r)dt^2 + \frac{1}{g(r)}dr^2 + \frac{r^2}{L^2}(dx^2+dy^2+dz^2)\;,
\end{align}
where the radial coordinate ranges from the black brane horizon
at $r=r_h$ to the AdS boundary limit of $r \to \infty$, the parameter $L$ is the AdS radius of curvature, and the functions $f(r)$, $g(r)$ must 
satisfy  $\lim_{r\to r_h} f(r)=g(r)= 0$ and $\lim_{r\to \infty} f(r)=g(r)= \frac{r^2}{L^2}$~. 

Metric fluctuations are represented by 
$\bar{g}_{\mu\nu} = g_{\mu\nu} + h_{\mu\nu}$~, where $g_{\mu\nu}$ is the AdS background metric of~\cref{ads-metric} and $h_{\mu\nu}$ is a small perturbation of the background metric (\emph{i.e}, a graviton). In this analysis, the gravitons will propagate along the $z$-coordinate and so
\begin{align}\label{graviton ansatz}
h_{ab} \;=\;  \phi(r) e^{i(\omega t - k z)}\;,
\end{align}
where $\phi(r)$ is some well-behaved function.

We assume weak coupling and consider a perturbative expansion of the Lagrangian in terms of $\epsilon = \frac{\ell_s^2}{L^2} \ll 1$ ($\ell_s$ is the string length) as appropriate for the case of large $N$ and large but relatively smaller $\lambda$. The expansion can be expressed schematically as
\begin{align}\label{full-lagrangian}
\mathcal{L}\; =\; \sqrt{-g} \frac{R}{16\pi G_5}(1 + \epsilon L^2 R + \epsilon^2 L^4 R^2 + \cdots + \epsilon^k L^{2k}R^k + \cdots)\;,
\end{align}
where we have suppressed indices (all terms are contractions of 4-index 
Riemann tensors as previously discussed) and $G_5$ is Newton's constant in five dimensions. For the sake of brevity, the Lagrangian can also be written as
\begin{align}\label{schem1-hole-in-1}
\mathcal{L} \;=\; \sum_{k=1} L_k\;,
\end{align}   
where $L_k \sim R^k$ is meant to describe a contraction of $k$ Riemann tensors. So that, when referring to (\emph{e.g}) six-derivative gravity, we really mean the Lagrangian $L_3 = \frac{\sqrt{-g}R}{16\pi G_5} [\epsilon^2 L^4 R^2]$. 

Gravitational scattering amplitudes must be holographically renormalized if we are to make a connection with the gauge--gravity duality. As shown in \cite{Brustein1}, this process eliminates the need to consider the radial derivatives of gravitons as these would lead to divergences in the amplitude even after renormalization. Such terms would then be discarded in the gauge theory after the standard techniques of holographic renormalization have been applied~\cite{Skenderis20}. Hence, for the current work, only the $t$ and $z$ derivatives of  gravitons are considered.

Let us now clarify what is meant by the ``high-momentum regime''. Our motivation is the idealized case of a kinematic region that is potentially accessible via experiment~\footnote{The viability of experimental accessibility hinges on having $\epsilon$ not too small. See \cite{Brustein3} for further details.} and that allows one to distinguish between the contributions of the different $L_k$'s in~\cref{schem1-hole-in-1}. The basic idea is that larger values of momenta can compensate for larger powers of $\epsilon$ as one probes theories which are higher order in numbers of derivatives. This is accomplished by insisting that a contribution to amplitudes of order $\epsilon^q$ always include $2q+2$ derivatives acting on gravitons. Other contributions are to be discarded. So that, if $L^2\epsilon \omega^2$ is not too small a number, the  surviving contributions should be prominent in the scattering profiles (and ultimately in the gauge-theory correlators). 

Let us explain further what classifies a value of momentum as ``high" in this context. Firstly, background derivatives go as $\nabla_r\nabla_r \sim \frac{1}{L^2}$~,  while graviton momenta go as $\nabla_t\nabla_t \sim \omega^2$ and $\nabla_z\nabla_z \sim k^2$~; therefore, one requirement is that $\omega^2,k^2 \gg \frac{1}{L^2}$~. Furthermore, to apply the tools of the gauge--gravity duality, the hydrodynamic regime must necessarily be in effect. This means that $\omega \ll T$ and, consequently,
\begin{align}
1 \;\ll\; L\omega \;\ll\; TL\;.
\end{align}  
This range is indeed viable because, according to the duality, $TL \sim \frac{r_h}{L} \gg 1$~\cite{Witten}. One then needs only to hope that $\epsilon (\omega L)^2$ is sufficiently large.

We also work in the radial gauge, which means that  $h_{ra} = 0$ for any $a$~. This gauge divides the gravitonal perturbations into three distinct sectors: scalar, vector and tensor~\cite{Policastro:radialgauge}. However, scalar modes do not contribute to scattering amplitudes in the high-momentum regime. This is because scalars need to be sourced and sources can be expected, on general grounds, to introduce an additional factor of $\epsilon$. Meanwhile, vector modes are analogous to the electromagnetic potential and, as such, can either be gauged away or require a source when appearing in gauge-invariant combinations. Hence, these modes can also be discounted. Other fields can only  couple to the gravitons through a stress-tensor, which will  invoke additional powers of $\epsilon$. This leaves only the tensor modes $h_{xy}$ as relevant in the high-momentum regime. 

Given that the number of derivatives acting on gravitons has been
maximized, the only other type of tensor-mode  amplitude  requiring suppression is one that includes $\Box h_{xy} = 0 + \mathcal{O}(\epsilon)$~, since this adds an extra factor of $\epsilon$. Note though that an expression like $g^{ab}g^{cd}\nabla_a \nabla_c h_{xy}$ would survive (with the free indices suitably contracted). We should also add that, to maintain general covariance, tensor modes must come in pairs (see below). And so, with the restriction to tensor modes, any
non-trivial scattering amplitude  will necessarily be even. 

Finally, we will perturbatively expand the metric determinant and contravariant metrics by adopting the conventions of~\cite{tHooft}. To briefly review, a 
perturbed metric such as 
\begin{align}
	\bar{g}_{\mu\nu} \;=\; g_{\mu\nu} + h_{\mu\nu} 
\end{align}
would have a contravariant counterpart of the form
\begin{align}\label{hooft1}
	\bar{g}^{\mu\nu} = g^{\mu\nu} - h^{\mu\nu} + h^\mu_\rho h^{\rho\nu} +\mathcal{O}(h^3)
\end{align}
and the metric determinant is then
\begin{align}\label{hooft2}
	\sqrt{-\bar{g}} \; =\; \sqrt{-g}\left[ 1 + \frac{1}{2}h^\mu_\mu - \frac{1}{4}h^\mu_\rho h^\rho_\mu + \frac{1}{8}(h^\mu_\mu)^2 + \mathcal{O}(h^3) \right]\;.
\end{align}

Since our interest is in the tensor modes $h_{xy}$~, only expressions with an even number of gravitons will survive in~\cref{hooft1,hooft2}, and undifferentiated gravitons will have to come from either 
\begin{align}
	\bar{g}^{xx} \;=\; g^{xx} + h^x_yh^{yx} + (h^x_yh^{yx})^2 + \cdots + (h^x_yh^{yx})^p 
\end{align}
or
\begin{align}
	\sqrt{-\bar{g}}\; =\; \sqrt{-g}\left[ 1 - \frac{1}{2}h^x_yh^y_x - \frac{1}{2^2 2!}(h^x_yh^y_x)^2 + \cdots + \Theta(p)(h^x_yh^y_x)^p \right]\;,
\end{align}
where
\begin{align}
	\Theta(p)\; =\; -\frac{\Gamma[p-\frac{1}{2}]}{2\sqrt{\pi}p!}\;, 
\;\;\;\text{   for } p \in \mathbb{Z}\;.
\end{align}
As for the differentiated gravitons, these will of course come from expansions of the Riemann tensors (see below).

\section{Multi-point amplitudes}
\subsection{Basis multi-point amplitudes}

We start here by introducing some shorthand notation for perturbations of the Riemann tensor,
\begin{align}
\delta^{(1)}\mathcal{R}_{abcd}\; &=\; \nabla_b \Gamma_{dac}(h) - \nabla_c \Gamma_{dab}(h)\;,  \label{d1} \\
\delta^{(2)}\mathcal{R}_{abcd}\; &= g^{ef}\;\left[\Gamma_{eac}(h)\Gamma_{fbd}(h)-\Gamma_{ead}(h)\Gamma_{fbc}(h)\right]\;,\label{d2}
\end{align}
where
\begin{align}
\Gamma_{abc}(h) \;=\; \frac{1}{2}\left( \nabla_b h_{ac} + \nabla_c h_{ab} - \nabla_a h_{bc} \right) \;.
\end{align}
Given that the high-momentum regime is in effect (so that the background contributions from any Riemann tensor can be ignored), then each Riemann tensor effectively contributes
\begin{align}\label{inversion}
R_{abcd}\;\to\;\mathcal{R}_{abcd} \;\equiv\; \delta^{(1)}\mathcal{R}_{abcd} + \delta^{(2)}\mathcal{R}_{abcd}\;.
\end{align}

Recalling that gravitons can only be differentiated with respect to $t$ and $z$,
one can observe that any perturbed Riemann tensor must have a specific arrangement of indices. Up to symmetries,~\footnote{In particular, $\mathcal{R}_{abcd} = \mathcal{R}_{cdab}$ and $\mathcal{R}_{abcd}=-\mathcal{R}_{abdc}$.}
 these would be  $\delta^{(1)}\mathcal{R}_{axby}$ for  $a,b=\{ t,z\}$
and $\delta^{(2)}\mathcal{R}_{cxdx} $  for   $c,d=\{ t,z,y\}$~, 
where  $x$ and $y$ are interchangeable. 

With the high-momentum regime in mind, we will define basis amplitudes as multi-point amplitudes for which \emph{all} of the included gravitons are differentiated. Each such basis amplitude then represents a different combination of $\delta^{(1)}\mathcal{R}$'s and $\delta^{(2)}\mathcal{R}$'s. Furthermore, in constructing a general $2n$-point amplitude, we will use the notation
 $\avg{h^{2n}}_{2p}$ to indicate a $2n$-point amplitude constructed from a $2p$-point basis amplitude ($n\geq p$). For instance, the basis amplitudes themselves are denoted by $\avg{h^{2p}}_{2p}$.

\subsection{Multi-point amplitudes from Riem$^3$ ($\epsilon^2$-order) gravity}\label{section-riem3gravity}
Six-derivative or Riem$^3$ gravity can be defined, up to gauge transformations, as 
\begin{align}\label{riem3}
L_3 \;=\; \sqrt{-g}\bigg[ \alpha \tensor{R}{_{abcd}}\tensor{R}{^{ab}_{mn}}\tensor{R}{^{mncd}} + \beta \tensor{R}{_{abcd}}\tensor{R}{_{mn}^{ad}}\tensor{R}{^{mncb}}\bigg]\;,
\end{align}
where $\frac{\epsilon^2 L^4}{16\pi G_5}$ has now been absorbed into the model-dependent constants $\alpha$ and $\beta$. 

In this theory, there two types of basis amplitudes;~\footnote{Here, we will be disregarding 2-point amplitudes because they are not really of interest from the viewpoint of someone discriminating between different theories.} the 4 and 6-point amplitudes. 
The basis 4-point amplitude has 6 derivatives and 4 gravitons; schematically it can written in terms of the previously introduced shorthand ({\em cf},~\cref{d1,d2}),
\begin{align}\label{d3}
\avg{h^4}_4 \;\sim\; 3(\alpha + \beta)\; \delta^{(1)}\mathcal{R}\delta^{(1)}\mathcal{R}\delta^{(2)}\mathcal{R}\;, 
\end{align}
where the tensor  indices have been suppressed  and the 3 
counts the number of ways of choosing one of  the  tensors to carry
two gravitons.

Similarly, the 6-point amplitude with 6 derivatives and 6 gravitons takes the schematic form
\begin{align}
\avg{h^6}_6 \;\sim\; (\alpha + \beta)\;\delta^{(2)}\mathcal{R}\delta^{(2)}\mathcal{R}\delta^{(2)}\mathcal{R}\;.
\end{align}

Now more explicitly, the 4-point amplitude of~\cref{d3}  can be expressed as
\begin{align}\label{d4}
\hspace*{-0.2in}\avg{h^4}_4 \;&=\; 3(\alpha + \beta) 2^3 \bigg[ \mathcal{R}_{txty}\mathcal{R}^{txty}\tensor{\mathcal{R}}{^{tx}_{tx}} + \mathcal{R}_{txzy}\mathcal{R}^{txzy}\tensor{\mathcal{R}}{^{tx}_{tx}} + \mathcal{R}_{txzy}\mathcal{R}^{zxzy}\tensor{\mathcal{R}}{^{tx}_{zx}} + \{ t \leftrightarrow z \}\bigg] \notag \\ \;&+\; \{x \leftrightarrow y \}\;,
\end{align}
where $\{ a \leftrightarrow b\}$ is shorthand for the  interchange of $a$ and $b$ in the preceding expression and the $2^3$ accounts for the symmetries of the Riemann tensors. Even more explicitly, in terms of gravitons (and with
the help of the equations in Subsection~2.1),
\begin{align}\label{base4}
\hspace*{-0.2in}\avg{h^4}_4 \;=\; -6(\alpha+\beta)\sqrt{-g}(g^{xx}g^{yy})^2 &\bigg[(\omega_1g^{tt}\omega_2 + k_1g^{zz}k_2)(\omega_1g^{tt}\omega_3 + k_1g^{zz}k_3) (\omega_2g^{tt}\omega_4 + k_2g^{zz}k_4 )  \bigg] \notag \\ &\times h_{xy}^{(1)}h_{xy}^{(2)}h_{xy}^{(3)}h_{xy}^{(4)}\;,
\end{align}
where each factor of $(\omega_i g^{tt}\omega_j + k_ig^{zz}k_j)$ is a product of the momenta for gravitons $h_{xy}^{(i)}$ and $h_{xy}^{(j)}$~, and the symmetrization of the gravitons is always implied. 

Using similar reasoning, one finds that the basis 6-point amplitude 
translates into
\begin{align}\label{base6}
\hspace*{-0.2in}\avg{h^6}_6\; =\; -\frac{3}{2}(\alpha+\beta)\sqrt{-g}(g^{xx}g^{yy})^3 &\bigg[(\omega_1g^{tt}\omega_2 + k_1g^{zz}k_2)(\omega_3g^{tt}\omega_4 + k_3g^{zz}k_4) (\omega_5g^{tt}\omega_6 + k_5g^{zz}k_6 )  \bigg] \notag \\ &\times h_{xy}^{(1)}h_{xy}^{(2)}h_{xy}^{(3)}h_{xy}^{(4)}h_{xy}^{(5)}h_{xy}^{(6)}\;.
\end{align}

The utility of the basis amplitudes in Eqs.~(\ref{base4}) and~(\ref{base6}) is that  one can use these to  construct $2n$-point amplitudes by drawing out additional  pairs of gravitons from the metric determinant  and the contravariant metrics $g^{xx}$, $g^{yy}$~. 
Let us begin with Eq.~(\ref{base4}) and
suppose that $p$ pairs are drawn  from the metric determinant and $n-2-p$ pairs from the four contravariant metrics. Recalling that the number of ways of drawing $q$ identical objects from $m$ distinct boxes is $\begin{pmatrix} q+m-1 \\ m - 1\end{pmatrix}$~,
 we then  have  that, for any $n\geq 2$~,
\begin{align}\label{base4n}
\avg{h^{2n}}_4 \;&=\; -6(\alpha+\beta)\begin{pmatrix} 2n \\ 4 \end{pmatrix} \sum_{p=0}^{n-2} \begin{pmatrix} n+1-p \\ 3 \end{pmatrix} \Theta(p)\sqrt{-g}(g^{xx}g^{yy})^2  \notag  \\ \;&\times\;\bigg[  \big( \omega_1g^{tt}\omega_2 + k_1g^{zz}k_2 \big) \notag  \big( \omega_1g^{tt}\omega_3 + k_1g^{zz}k_3 \big)\big( \omega_2g^{tt}\omega_4 + k_2g^{zz}k_4 \big)   \bigg] \notag \\ \;&\times\; h^{(1)}_{xy}h^{(2)}_{xy}h^{(3)}_{xy}h^{(4)}_{xy}\prod_{j=3}^{n}\bigg[ (h^x_y)^{(2j-1)}(h_x^y)^{(2j)} \bigg]\;,
\end{align}  
where the combinatorial factor before the summation represents the number of ways of choosing the  four differentiated gravitons from the total of $2n$ and
 the summation itself accounts for all  possible ways of  drawing gravitons 
from the contravariant metrics and  metric determinant. 

In similar fashion,  $2n$-point amplitudes
can be constructed    from the basis 6-point amplitude of~\cref{base6}
for any $n \geq 3$~. The result of this is
\begin{align}\label{base6n}
\avg{h^{2n}}_6 \;&=\; -\frac{3}{2}(\alpha+\beta) \begin{pmatrix} 2n\\ 6 \end{pmatrix}\sum_{q=0}^{n-3}\begin{pmatrix} n-q+2 \\ 5 \end{pmatrix} \Theta(q) \sqrt{-g}(g^{xx}g^{yy})^3 \notag \\ \;&\times\; \bigg[  \big( \omega_1g^{tt}\omega_2 + k_1g^{zz}k_2 \big) \big( \omega_3g^{tt}\omega_4 + k_3g^{zz}k_4 \big)  \big( \omega_5g^{tt}\omega_6 + k_5g^{zz}k_6 \big) \bigg] \notag \\ \;& \times\;  h^{(1)}_{xy}h^{(2)}_{xy}h^{(3)}_{xy}h^{(4)}_{xy}h^{(5)}_{xy}h^{(6)}_{xy}\prod_{k=4}^{n} \bigg[ (h^x_y)^{(2k-1)}(h_x^y)^{(2k)} \bigg] \;.
\end{align}

As~\cref{base4n,base6n} now make clear, the 4 and 6-point amplitudes of~\cref{base4,base6} form the basis for the $2n$-point amplitudes of Riem$^3$ gravity 
in the high-momentum regime. 
The complete $2n$-point amplitude for Riem$^3$ gravity is then given by the linear combination
\begin{align}
\avg{h^{2n}}_{\text{Riem}^3} \;=\; \avg{h^{2n}}_4 + \avg{h^{2n}}_6\;.
\end{align}

\subsection{Multi-point amplitudes from Riem$^4$ ($\epsilon^3$-order) gravity}\label{section-riem4gravity}
Eight-derivative or Riem$^4$ gravity is expressible, up to gauge transformations, as
\begin{align}\label{riem4}
\hspace*{-0.2in}L_4 \;&=\; \bigg[ \alpha \tensor{R}{_{abcd}}\tensor{R}{^{abmn}}\tensor{R}{_{mn}^{pq}}\tensor{R}{_{pq}^{cd}} + \beta \tensor{R}{_{abcd}}\tensor{R}{^{ab}_{qp}}\tensor{R}{^{mndq}}\tensor{R}{_{mn}^{cp}} + \gamma \tensor{R}{_{abcd}}\tensor{R}{^{pd}_{mn}}\tensor{R}{^{abcq}}\tensor{R}{_{pq}^{mn}} \notag \\ \;&+\; \mu \tensor{R}{_{abcd}}\tensor{R}{_{mnpq}}\tensor{R}{^{ancq}}\tensor{R}{^{mbpd}} + \nu \tensor{R}{_{abcd}}\tensor{R}{^{na}_{pq}}\tensor{R}{^{mbpd}}\tensor{R}{_{mn}^{qc}} + \rho \tensor{R}{^{abcd}}\tensor{R}{_{abcd}}\tensor{R}{^{mnpq}}\tensor{R}{_{mnpq}} \bigg]\;.
\end{align}
Here, like before, $\frac{\epsilon^3 L^6}{16\pi G_5}$ has been absorbed into the model-dependent constants.

With the very same reasoning as in the previous section, we can call on~\cref{riem4} to construct three types of basis amplitudes; the 4, 6 and 8-point amplitudes. In terms of the Riemann tensor in~\cref{inversion}, these can be schematically written as
\begin{align}
\avg{h^4}_4\; &\sim\; \left(\alpha + \beta + \gamma + \mu + \nu + \rho\right) \delta^{(1)}\mathcal{R}\delta^{(1)}\mathcal{R}\delta^{(1)}\mathcal{R}\delta^{(1)}\mathcal{R}  \;, \label{scem-riem4-4pf} \\ 
\avg{h^6}_6 \;&\sim\; 6(\alpha + \beta + \gamma + \mu + \nu + \frac{1}{3}\rho) \delta^{(1)}\mathcal{R}\delta^{(1)}\mathcal{R}\delta^{(2)}\mathcal{R}\delta^{(2)}\mathcal{R}\; , \label{scem-riem4-6pf} \\
\avg{h^8}_8 \;&\sim\; (\alpha + \beta + \gamma + \mu + \nu + \rho) \delta^{(2)}\mathcal{R}\delta^{(2)}\mathcal{R}\delta^{(2)}\mathcal{R}\delta^{(2)}\mathcal{R}\;. \label{scem-riem4-8pf} 
\end{align}

Expanding~\cref{scem-riem4-4pf}, one finds that
\begin{align}\label{r-4pf-final}
\hspace*{-0.2in}\avg{h^4}_4\; =&\; 4 \left(\alpha + \beta + \gamma + \mu + \nu + \rho\right) \sqrt{-g}(g^{xx}g^{yy})^2 \bigg[(\omega_1 g^{tt} \omega_2 + k_1g^{zz}k_2 ) (\omega_2 g^{tt} \omega_3 + k_2g^{zz}k_3 )\notag \\ \;&\times\; (\omega_3 g^{tt} \omega_4 + k_3g^{zz}k_4 )(\omega_1 g^{tt} \omega_4 + k_1g^{zz}k_4 )  \bigg] h^{(1)}_{xy}h^{(2)}_{xy}h^{(3)}_{xy}h^{(4)}_{xy}\;,
\end{align}
which can  then be used to construct a $2n$-point amplitude for any $n \geq 2$,
\begin{align}\label{4-4-2n}
\hspace*{-0.2in}\avg{h^{2n}}_4 \;&=\;\left(\alpha + \beta + \gamma + \mu + \nu + \rho\right)\begin{pmatrix} 2n \\ 4 \end{pmatrix} \sum_{p=0}^{n-2} \begin{pmatrix} n+1-p \\ 3 \end{pmatrix} \Theta(p)\sqrt{-g}(g^{xx}g^{yy})^2   \notag \\ \;& \times\; \bigg[ (\omega_1 g^{tt} \omega_2 + k_1g^{zz}k_2 )(\omega_2 g^{tt} \omega_3 + k_2g^{zz}k_3 ) (\omega_3 g^{tt} \omega_4 + k_3g^{zz}k_4 )(\omega_1 g^{tt} \omega_4 + k_1g^{zz}k_4 )   \bigg] \notag \\\; &\times\; h^{(1)}_{xy}h^{(2)}_{xy}h^{(3)}_{xy}h^{(4)}_{xy} \prod_{j=3}^{n}\bigg[ (h^x_y)^{(2j-1)}(h_x^y)^{(2j)} \bigg]\;,
\end{align}
where the combinatoric factors (here and below) are handled similarly to those in~\cref{base4n}.

As for the basis 6-point amplitude, this goes as
\begin{align}\label{6pf-full-pre}
\avg{h^6}_6 \;=&\; \bigg(4\rho+6 (\alpha + \beta + \gamma + \mu + \nu)\bigg)\sqrt{-g}(g^{xx}g^{yy})^3 \notag \\\; & \times\; \big[ (\omega_1g^{tt}\omega_2 + k_1 g^{zz} k_2)^2(\omega_3g^{tt}\omega_4 + k_3 g^{zz} k_4)(\omega_5g^{tt}\omega_6 + k_5 g^{zz} k_6) \big] \notag \\\; & \times\; h^{(1)}_{xy}h^{(2)}_{xy}h^{(3)}_{xy}h^{(4)}_{xy}h^{(5)}_{xy}h^{(6)}_{xy}\;,
\end{align}
from which one can construct a $2n$-point amplitude for any $n \geq 3$~,
\begin{align}\label{4-6-2n}
\hspace*{-0.2in}\avg{h^{2n}}_6 \;=&\;  \bigg(4\rho+6(\alpha + \beta + \gamma + \mu + \nu)\bigg)\begin{pmatrix} 2n\\ 6 \end{pmatrix}\sum_{q=0}^{n-3}\begin{pmatrix} n-q+2 \\ 5 \end{pmatrix} \Theta(q) \sqrt{-g}(g^{xx}g^{yy})^3  \notag \\ \;& \times\; \big[ (\omega_1g^{tt}\omega_2 + k_1 g^{zz} k_2)^2 (\omega_3g^{tt}\omega_4 + k_3 g^{zz} k_4)(\omega_5g^{tt}\omega_6 + k_5 g^{zz} k_6) \big] \notag \\\; & \times\; h^{(1)}_{xy}h^{(2)}_{xy}h^{(3)}_{xy}h^{(4)}_{xy}h^{(5)}_{xy}h^{(6)}_{xy}\prod_{k=4}^n\bigg[ (h^x_y)^{(2k-1)}(h_x^y)^{(2k)} \bigg]\; .
\end{align}

Finally, the basis 8-point amplitude is of the form
\begin{align}\label{riem4-basis-8pf}
\hspace*{-0.2in}\avg{h^8}_8 \;&=\; \bigg(\rho+\frac{1}{2}(\alpha + \beta + \gamma + \mu + \nu)\bigg)(g^{xx}g^{yy})^4 (\omega_1 g^{tt}\omega_2 + k_1 g^{zz}k_2)(\omega_3 g^{tt}\omega_4 + k_3 g^{zz}k_4)\notag \\ \;& \times\;(\omega_5 g^{tt}\omega_6 + k_5 g^{zz}k_6) (\omega_7 g^{tt}\omega_8 + k_7 g^{zz}k_8)  h^{(1)}_{xy}h^{(2)}_{xy}h^{(3)}_{xy}h^{(4)}_{xy}h^{(5)}_{xy}h^{(6)}_{xy}h^{(7)}_{xy}h^{(8)}_{xy}\;,
\end{align}
and the corresponding $2n$-point amplitude for any $n \geq 4$ is then
\begin{align}\label{4-8-2n}
\hspace*{-0.2in}\avg{h^{2n}}_8 \;=&\; \bigg(\rho+\frac{1}{2}(\alpha + \beta + \gamma + \mu + \nu)\bigg) \begin{pmatrix} 2n \\ 8 \end{pmatrix} \sum_{r=0}^{n-4}\begin{pmatrix} n-r+3 \\ 7 \end{pmatrix} \Theta(r) \sqrt{-g}(g^{xx}g^{yy})^4  \notag \\\; & \times\; (\omega_1 g^{tt}\omega_2 + k_1 g^{zz}k_2)(\omega_3 g^{tt}\omega_4 + k_3 g^{zz}k_4) (\omega_5 g^{tt}\omega_6 + k_5 g^{zz}k_6)(\omega_7 g^{tt}\omega_8 + k_7 g^{zz}k_8) \notag \\\; &\times\;  h^{(1)}_{xy}h^{(2)}_{xy}h^{(3)}_{xy}h^{(4)}_{xy}h^{(5)}_{xy}h^{(6)}_{xy}h^{(7)}_{xy}h^{(8)}_{xy}\prod_{l=5}^n\bigg[ (h^x_y)^{(2l-1)}(h_x^y)^{(2l)} \bigg]\;.
\end{align}

Like before, the complete $2n$-point amplitude for the  Riem$^4$ theory in the high-momentum regime is a linear combination of the basis amplitudes in~\cref{4-4-2n,4-6-2n,4-8-2n}. That is,
\begin{align}
\avg{h^{2n}}_{\text{Riem}^4}\; =\; \avg{h^{2n}}_4 + \avg{h^{2n}}_6 + \avg{h^{2n}}_8
\;.
\end{align}

\section{Multi-Point amplitudes from Riem$^q$ ($\epsilon^{q-1}$-order) gravity}\label{section-riemqgravity}
In this section, we will find the basic  form of the $2n$-point amplitude 
in the high-momentum regime when the most general type of gravitational theory is considered. Namely, one 
whose Lagrangian is composed  of $q$ contracted Riemann tensors for arbitrary $q$. 

This task appears to be quite arduous, as  one would expect 
that the number of gauge-invariant terms in the Lagrangian
grows exponentially with $q$.~\footnote{It
is amusing to note that the number of gauge-invariant terms grows 
exactly as   $(q-1)!$ for $q \leq 4$~, which then  grows roughly
as $e^{q}$ for large $q$.} On the other hand, as shown in 
Subsection~2.3, each of the six invariants makes essentially
the same contribution to  any one of the three basis amplitudes
 ({\em cf},~\cref{r-4pf-final,6pf-full-pre,riem4-basis-8pf}) ---
although the different basis amplitudes will indeed have different forms. It is not difficult to convince oneself that the relative simplicity of the high-momentum regime is enough to ensure that these trends will   persist to higher orders
in $q$.

Since any single graviton can carry zero, one or two derivatives, there are many ways to obtain a $2n$-point amplitude from a Riem$^q$ theory depending on the size and parity of $q$. Let us suppose that this is a $q$-odd theory; then the set of basis amplitudes ({\em i.e.}, those with only differentiated gravitons) is the set $P_q=\{ \avg{h^{q+1}}_{q+1},\avg{h^{q+3}}_{q+3},\dots,\avg{h^{2q}}_{2q} \}$ with cardinality $\frac{q+1}{2}$\;. The reasoning here is that, for an odd value of $q$, $\avg{h^{q+1}}_{q+1}$ has the maximum possible number of gravitons carrying two derivatives with the remainder carrying one, whereas $\avg{h^{2q}}_{2q}$ has all gravitons carrying a single derivative.    

To elaborate further, let us consider the following arrangements of $q$ contracted Riemann tensors (while keeping in mind~\cref{d1,d2} and that the gravitons come in pairs): 
\begin{align}
	\avg{h^{q+1}}_{q+1}\; &\sim\; \underbrace{\delta^{(1)}\mathcal{R}\delta^{(1)}\mathcal{R} \cdots \delta^{(1)}\mathcal{R}\delta^{(2)}\mathcal{R}}_{q\text{ products of Riemann tensors}} \; \\ & \  \ \text{a single}\; \delta^{(2)}\mathcal{R} \text{ term and an even number of }\delta^{(1)}\mathcal{R}\text{ terms,} \notag \\
	\avg{h^{q+3}}_{q+3} \;& \sim\; \underbrace{\delta^{(1)}\mathcal{R}\delta^{(1)}\mathcal{R} \cdots \delta^{(2)}\mathcal{R}\delta^{(2)}\mathcal{R}\delta^{(2)}\mathcal{R}}_{q\text{ products of Riemann tensors}}\; \\& \ \ \text{ three } \delta^{(2)}\mathcal{R} \text{ terms and an even number of }\delta^{(1)}\mathcal{R}\text{ terms,} \notag \\
	& \vdots \qquad \vdots \qquad \vdots \qquad \vdots \qquad \qquad \qquad \qquad \qquad \vdots \notag \\
	\avg{h^{2q}}_{2q} \;&\sim\;  \underbrace{\delta^{(2)}\mathcal{R}\delta^{(2)}\mathcal{R} \cdots \delta^{(2)}\mathcal{R}\delta^{(2)}\mathcal{R}}_{q \text{ products of Riemann tensors}}\;  \\ & \ \ \text{ all } \delta^{(2)}\mathcal{R} \text{ terms.} \notag
\end{align}
A similar argument can be used for $q$-even theories; in which case, the set of basis amplitudes is $Q_q = \{\avg{h^{q}}_q,\avg{h^{q+2}}_{q+2},\dots,\avg{h^{2q}}_{2q} \}$ with cardinality $\frac{q+2}{2}$~. In both cases, the key point is that there should either be zero or an even number of $\delta^{(1)}\mathcal{R}$'s 
in any basis amplitude. 

The next step involves reformulating the different basis amplitudes in terms of gravitons rather than Riemann tensors. It is clear that each such basis amplitude will be a polynomial in $\omega$'s and $k$'s of degree $2q$. Since either one or two derivatives can act on a graviton, each term in the polynomial must contain one of $\omega_{i},\;k_{i},\;\omega_{i}k_{i},\; \omega_{i}^2,\; k_{i}^2$ for each graviton $h^{(i)}_{xy}$. Then as long as $ q \leq 2p < 2q$ for some $p \in \mathbb{N}^{+}$~, a basis amplitude for Riem$^q$ can be expressed somewhat schematically as~\footnote{This form assumes that the gravitons can be freely labeled.}
\begin{align}
\avg{h^{2p}}_{2p} \;&=\; \mathcal{A}_{q} 2^{q+2-2p}\sqrt{-g}(g^{xx}g^{yy})^p \underbrace{C_{(1,2)}C_{(3,4)} \cdots C_{(2p-1,2p)} C_{(1,2p)} \cdots }_{ q \text{ contractions of derivatives} }  \notag \\ \;& \times\; h_{xy}^{(1)}h_{xy}^{(2)}h_{xy}^{(3)}h_{xy}^{(4)} \cdots h_{xy}^{(2p-1)}h_{xy}^{(2p)}\;,
\end{align} 
where $C_{(i,j)} \equiv K_i \cdot K_{ j}  = (\omega_i g^{tt} \omega_j + k_ig^{zz}k_j)$ with $K_i=(\omega_i,0,0,0,k_i)$ and $\mathcal{A}_q \sim \mathcal{O}(e^q)$ is a numerical coefficient. For $2p = 2q$~, the basis amplitude should rather
be  written as
\begin{align}
\avg{h^{2q}}_{2p}\; &=\; \mathcal{A}_{q} 2^{2-q}\sqrt{-g}(g^{xx}g^{yy})^q \underbrace{C_{(1,2)}C_{(3,4)} \cdots C_{(2q-1,2q)}}_{ q \text{ contractions of derivatives} } \notag \\ \;& \times\; h_{xy}^{(1)}h_{xy}^{(2)}h_{xy}^{(3)}h_{xy}^{(4)} \cdots h_{xy}^{(2q-1)}h_{xy}^{(2q)}\;.
\end{align}

Now each basis amplitude $\avg{h^{2p}}_{2p}$ --- for which $q \leq 2p \leq 2q$ if 
$q$ is even or $q+1 \leq 2p \leq 2q$ if $q$ is odd --- will contribute to the $2n$-point amplitude in accordance with
\begin{align}\label{2p-2n}
	\avg{h^{2n}}_{2p} \;&=\; \mathcal{A}_q 2^{q+2-2p} \begin{pmatrix} 2n \\ 2p \end{pmatrix} \sum_{r=0}^{n-p}\begin{pmatrix} n+p-r+1 \\ 2p -1 \end{pmatrix} \Theta(r) \sqrt{-g}(g^{xx}g^{yy})^p \notag \\ \;& \times\; \underbrace{C_{(1,2)}C_{(3,4)} \cdots C_{(2p-1,2p)} C_{(1,2p)} \cdots }_{ q \text{ contractions} } \notag \\ 
\; &\times\;  h^{(1)}_{xy}h^{(2)}_{xy} \cdots h^{(2p-1)}_{xy}h^{(2p)}_{xy}\prod_{m=p+1}^n\bigg[  (h^x_y)^{(2m-1)}(h_x^y)^{(2m)} \bigg]\;,
\end{align} 
for all $n \geq p $~, and the combinatoric factors follow the same logic as in
the analysis from Section~2.

 Finally, we can use~\cref{2p-2n} to express the complete $2n$-point for any Riem$^q$ theory gravity as the sum of contributions from the   various basis amplitudes,
\begin{align}
\hspace*{-0.2in}\avg{h^{2n}}_{\text{Riem}^q} \;=\; \left\{ \begin{array}{lr}
	\sum_{m=0}^{\frac{q}{2}} \avg{h^{2n}}_{2m+q}\;\; \text{  with $q$ even and } 2n \geq 2m+q \text{ for every }m\;, \\ \\
	\sum_{m=0}^{\frac{q-1}{2}} \avg{h^{2n}}_{2m+q+1}\;\; \text{  with $q$ odd and } 2n \geq 2m+q+1 \text{ for every }m\;.	
	\end{array} \right.
\end{align}
It should again be  emphasized that  the validity of these results depends upon the restriction to the  high-momentum regime.

\section{Scattering properties of $2n$-point amplitudes}\label{section-mandel}
Our next order of business is to look at the scattering properties of some
of these amplitudes. We will restrict to the cases with four gravitons (all of which are differentiated) as then the results can be expressed directly in terms of the familiar Mandelstam variables. However, it can be expected that the same basic theme ---  each theory carrying its own characteristic signature for scattering experiments in the high-momentum regime --- will persist to more complicated scenarios. 

Let us  begin with the $4$-point amplitude of Riem$^3$ gravity as depicted 
in Subsection~2.2. It is convenient to express the products of momenta in terms of the condensed notation $K_i \cdot K_j = (\omega_i g^{tt} \omega_j + k_i g^{zz} k_j)$~;
in which case,
\begin{align}\label{riem3-4pn-angular1}
\avg{h^4} \;=\; -6(\alpha + \beta)\sqrt{-g}(g^{xx}g^{yy})^2 K_{(1}\cdot K_2 \ K_3\cdot K_4 \ K_1\cdot K_{4)} h_{xy}^{(1)}h_{xy}^{(2)}h_{xy}^{(3)}h_{xy}^{(4)}\;.
\end{align}

It should be noted that the symmetrized product of momenta in~\cref{riem3-4pn-angular1} is really a shorthand for the  symmetrization of  all possible products in a particular way, 
\begin{align}\label{symmetrizing}
\hspace*{-0.2in}K_{(1}\cdot K_2 \ K_3\cdot K_4 \ K_1\cdot K_{4)}  \;&\to\; K_{(1}\cdot K_2 \ K_3\cdot K_4 \ K_1\cdot K_{4)} \;+\; K_{(1}\cdot K_2 \ K_3\cdot K_4 \ K_1\cdot K_{2)} \notag \\ \;&+\; K_{(1}\cdot K_2 \ K_3\cdot K_4 \ K_1\cdot K_{3)}\; +\; K_{(1}\cdot K_2 \ K_3\cdot K_4 \ K_2\cdot K_{3)} \notag \\\; &+\; K_{(1}\cdot K_2 \ K_3\cdot K_4 \ K_2\cdot K_{4)}\; +\; K_{(1}\cdot K_2 \ K_3\cdot K_4 \ K_3\cdot K_{4)}\;.     
\end{align}
To better appreciate~\cref{symmetrizing}, one can take note that each
of the momenta has an equal opportunity of appearing either once or twice  in any given permutation of the four gravitons. 

Let us now recall the Mandelstam variables,
\begin{align}
s\; &=\;  2K_1 \cdot K_2 = 2K_3 \cdot K_4\;, \\
t \;&=\; -2K_1 \cdot K_3 = -2K_2 \cdot K_4\;, \\
u \;&=\; -2K_1 \cdot K_4 = -2K_2 \cdot K_3\;,  
\end{align}
with $s+t+u=0$ for this case of massless particles. In terms of its dependency on the Mandelstam variables,~\cref{riem3-4pn-angular1} can be expressed as
\begin{align}\label{riem3-mandel}
\avg{h^4}_{\text{Riem}^3} \;\propto\; (s-t-u)(s^2+t^2+u^2) h_{xy}^{(1)}h_{xy}^{(2)}h_{xy}^{(3)}h_{xy}^{(4)}\;.
\end{align}

In the interest of making a connection with the gauge theory, it should be emphasized that \cref{riem3-mandel} is only valid at the AdS boundary where $g_{zz}=|g_{tt}|$ holds true. But, since the idea is to translate these expressions into statements in the dual gauge theory (as in~\cite{Brustein2}), the boundary limit is sufficient. Indeed, we expect to observe a related signature for the stress--energy correlators in the gauge theory. This is because of the observation in~\cite{Brustein1} that the amplitudes which survive in the high-momentum regime are mostly unaffected by the process of holographic renormalization.

We can similarly apply this process to the 4-point amplitude from Riem$^4$ gravity, which leads to
\begin{align}\label{rie4-mandel}
\avg{h^4}_{\text{Riem}^4}\; \propto\; (s^4+t^4+u^4+s^2t^2+s^2u^2 + u^2t^2) h_{xy}^{(1)}h_{xy}^{(2)}h_{xy}^{(3)}h_{xy}^{(4)}\;.
\end{align}

In view of~\cref{riem3-mandel,rie4-mandel}, two conclusions immediately follow: The first is that the two theories have very distinct scattering signatures and the second is that $s,\;t$ and $u$ appear democratically in both cases. The latter is a consequence of the high-momentum regime favoring no particular scattering channel. Meaning that, in general, this is also a regime of large-angle scattering.

\section{Conclusion}
In this paper, we have computed graviton multi-point scattering amplitudes for higher-derivative theories in an AdS black brane background. All computations were carried out in the so-called high-momentum regime as was first introduced in~\cite{Brustein1}. This regime allows for higher-curvature corrections to contribute significantly to higher-point amplitudes provided that $s\epsilon \approxless 1$, where $\epsilon$ is the perturbative ($\alpha^\prime$ or Regge slope) expansion parameter. Along with explicit calculations for six and eight-derivative theories, we were able to generalize the formalism to higher-derivative gravity of arbitrary order. A critical element of this generalization was the construction of a certain class of basis amplitudes. 

We proceeded to demonstrate the scattering properties of Riem$^3$ and Riem$^4$ gravity in terms of the Mandelstam variables $s$, $t$ and $u$ by using their respective 4-point amplitudes. Our expectation is that this procedure can be generalized to higher-point scattering amplitudes and higher-derivative theories with some amount of work.

The graviton multi-point amplitudes in this paper should correspond to stress--tensor correlators in the gauge theory. This means, for instance, that the stress--energy tensor 4-point correlators should include, in addition to the distinct signatures of Einstein gravity and four-derivative gravity~\cite{Brustein1,Brustein2}, those of six and eight-derivative gravity as depicted in~\cref{riem3-mandel,rie4-mandel}. Note, however, that, to make contact with actual experiments, it is the connected functions in the gauge theory that are required, whereas the amplitudes in this paper would correspond to 1PI functions. This point will be addressed at a later time~\cite{Shawa}.

  \section*{Acknowledgments}
                The research of AJMM received support from an NRF Incentive Funding Grant 85353 and  NRF Competitive Programme Grant 93595.  MMWS is supported by an NRF bursary through  Competitive Programme Grant 93595 and a Henderson Scholarship from Rhodes University.

\newpage

\bibliographystyle{unsrtnat}



	
\end{document}